\documentclass[aps,prl,twocolumn,preprintnumbers,superscriptaddress,dblfloatfix,nofootinbib]{revtex4-1}
\usepackage[utf8]{inputenc}
\usepackage{lmodern}
\usepackage{amsmath,amssymb}
\usepackage{graphicx}
\usepackage{url}
\usepackage{color}
\usepackage{enumitem}
\usepackage{subfigure}
\usepackage[dvipsnames]{xcolor}
\usepackage[colorlinks=true,breaklinks=true]{hyperref}
\hypersetup{allcolors=[rgb]{0.0 0.0 0.6},linkcolor=[rgb]{0.75 0.05 0.05}}
\usepackage{bm}
\usepackage{epsfig}
\usepackage{amsmath}
\usepackage{amssymb}
\usepackage{slashed}
\usepackage{color}
\usepackage{accents}
\usepackage[dvipsnames]{xcolor}
\usepackage[colorlinks=true,breaklinks=true]{hyperref}
\hypersetup{allcolors=[rgb]{0.0 0.0 0.6},linkcolor=[rgb]{0.75 0.05 0.05}}

\renewcommand\[{\left[}

\newcommand{\exclude}[1]{}

\begin{document}
\preprint{IPMU21-0038}

\title{Spins of primordial black holes formed in different cosmological scenarios}

\author{Marcos M.  Flores} 
\affiliation{Department of Physics and Astronomy, University of California, Los Angeles \\ Los Angeles, California, 90095-1547, USA} 
\author{Alexander Kusenko} 
\affiliation{Department of Physics and Astronomy, University of California, Los Angeles \\ Los Angeles, California, 90095-1547, USA}
\affiliation{
Kavli Institute for the Physics and Mathematics of the Universe (WPI),The University of Tokyo Institutes for Advanced Study, The University of Tokyo, Kashiwa, Chiba 277-8583, Japan
}

\date{\today}

\begin{abstract}
Primordial black holes (PBHs) could account for all or part of dark matter, as well as for some LIGO events. We discuss the spins of primordial black holes produced in different cosmological scenarios, with the emphasis on recently discovered possibilities.  PBHs produced as a horizon-size collapse of density perturbations are known to have very small spins.  In contrast, PBHs resulting from assembly of matterlike objects (particles, Q-balls, oscillons, etc.) can have large or small spins depending on their formation history and the efficiency of radiative cooling.  We show that scalar radiation can remove the angular momentum very efficiently, leading to slowly rotating PBHs in those scenarios for which the radiative cooling is important.  Gravitational waves astronomy offers an opportunity to determine the spins of black holes, opening a new window on the early Universe if, indeed, some black holes have primordial origin.

\end{abstract}
\maketitle


It is an open question whether black holes could have been created in the early Universe.  These primordial black holes (PBHs) can account for all or part of dark matter (DM) ~\cite{Zeldovich:1967,Hawking:1971ei,Carr:1974nx,Carr:1975qj,Khlopov:1980mg,Khlopov:1985jw,Yokoyama:1995ex,GarciaBellido:1996qt,Kawasaki:1997ju,Green:2004wb,Khlopov:2008qy,Carr:2009jm,Frampton:2010sw,Kawasaki:2016pql,Carr:2016drx,Inomata:2016rbd,Pi:2017gih,Inomata:2017okj,Garcia-Bellido:2017aan,Georg:2017mqk,Inomata:2017vxo,Kocsis:2017yty,Ando:2017veq,Cotner:2016cvr,Cotner:2017tir,Cotner:2018vug,Sasaki:2018dmp,Carr:2018rid,Banik:2018tyb,1939PCPS...35..405H,Kawasaki:2018daf,Cotner:2019ykd,Hasegawa:2018yuy,Inman:2019wvr,Escriva:2019phb,Germani:2019zez,Kawasaki:2019hvt,Kawasaki:2019iis,Kusenko:2020pcg,deFreitasPacheco:2020wdg,Flores:2020drq,Takhistov:2020vxs,Inomata:2020xad,Domenech:2021uyx,Domenech:2021wkk,Baker:2021nyl,Gross:2021qgx,Kawana:1866357,Kawasaki:2021zir,Cheek:2021odj}. 
Furthermore, PBHs can be responsible for the gravitational events detected by LIGO~\cite{Abbott:2016blz,Abbott:2016nmj,Abbott:2017vtc,Clesse:2016vqa,Bird:2016dcv,Sasaki:2016jop}
and can seed supermassive black holes \cite{Bean:2002kx,Kawasaki:2012kn,Clesse:2015wea}.

A number of different scenarios have been proposed for the formation of PBHs. For the purposes of our discussion, one can classify them into three classes: 
(i) collapse of horizon-size overdensities ~(e.g., Refs. \cite{Carr:1974nx,Carr:1975qj,Yokoyama:1995ex,GarciaBellido:1996qt,Kawasaki:1997ju,Deng:2017uwc,Kusenko:2020pcg}); (ii) subhorizon assembly of massive objects (such as heavy particles, supersymmetric Q-balls, or oscillons, e.g., ~\cite{Khlopov:1980mg,Khlopov:1985jw,Polnarev:1986bi,Cotner:2016cvr,Cotner:2017tir,Cotner:2018vug,Cotner:2019ykd}) (iii) subhorizon assembly of massive particles in the presence of radiative cooling due to Yukawa or other long-range forces~\cite{Flores:2020drq}.  These classes can be roughly characterized by the fraction of the horizon mass that is trapped in an individual black hole $\zeta = M_{\rm PBH}/M_{\rm H, f} $.  Inflationary models (i) result in PBH masses of the order of the horizon mass at the time of formation: $\zeta\sim 1$.  The models in the second class (ii) usually produce PBHs with $\zeta \sim 0.01-0.1$~\cite{Cotner:2019ykd}.  Finally, the models with radiative cooling deal with halos that can collapse into black holes at times that are much later than the halo formation time, leading to $\zeta \ll 1$, with no meaningful lower bound. We note that our classification does not refer to whether the Universe is matter or radiation dominated at the time of PBH formation, although most of the literature on case (i) deals with a radiation dominated universe, while models (ii) typically invoke an intermediate matter-dominated epoch, and scenarios (iii) can work in either radiation or matter dominated epochs. 

{\it In case (i)}, the near horizon size collapse of perturbations ($\zeta\sim 1$) leads to formation of a population of low-spin PBHs~\cite{Chiba:2017SpinDist, Luca:2019InitSpin, Mirbabayi:2019uph, Harada:2021Spins,Jaraba:2021ces}. 

{\it In case (ii)}, $\zeta < 1$ black holes are assembled from individual particles~\cite{Khlopov:1980mg} or fragments~\cite{Cotner:2016cvr,Cotner:2017tir,Cotner:2018vug,Cotner:2019ykd} on the subhorizon scales (which tends to be be relatively close to the horizon scale for a successful collapse~\cite{Cotner:2019ykd}).  A viable class of scenarios is based on fragmentation of a primordial scalar field (a supersymmetric flat direction, an inflaton, or any other scalar that has self-interactions causing an instability~\cite{Khlopov:1985jw,Kusenko:1997si}).  Since the lumps are heavy and few, their distribution is subject to relatively large fluctuations.  A positive fluctuation in the density of such scalar lumps can become a PBH if the configuration is sufficiently spherical and carries a small angular momentum~\cite{Polnarev:1986bi,Cotner:2019ykd}.  The positive density fluctuations are rare, and the additional selection by the angular momentum makes the formation of PBH strongly suppressed.  Nevertheless, the resulting mass density can be sufficient to explain dark matter or LIGO data~\cite{Cotner:2016cvr,Cotner:2016dhw,Cotner:2017tir,Cotner:2018vug,Cotner:2019ykd}.

Let us consider a subset of overdensities larger than some critical value needed for the collapse under the action of gravity~\cite{Carr:1974nx,Carr:1975qj,Khlopov:1985jw,Polnarev:1986bi,Cotner:2016cvr,Cotner:2016dhw,Cotner:2017tir,Cotner:2018vug,Cotner:2019ykd}. For definiteness, we will consider a scenario involving an intermediate matter-dominated era~\cite{Cotner:2019ykd}. The formation of a PBH requires that the Kerr spin parameter $a_{ * }\equiv J/GM_{\rm PBH}^2$ of the system is less than or equal to one. In the absence of angular momentum dissipation, halos with $a_{*} > 1$ are unable to collapse into black holes.

This condition is very restrictive as demonstrated in~\cite{Harada:2017Spins}. The distribution for the spin parameter depends on the variance of density fluctuations at horizon entry $\sigma_H$. When $\sigma_H \lesssim 0.04$, the distribution $f(a_{*})$ is a monotonically rising function for $0 < a_{*} < 1$. Therefore, the subset of black holes satisfying $a_* < 1$ is dominated by configurations with $a_*\approx 1$, and thus nearly extremal. Alternatively, when $\sigma_H \gtrsim 0.04$ the average spin parameter $\bar{a}_*$ is between zero and one. If one assumes that power spectrum of the primordial comoving curvature perturbations is scale invariant, $f_{\rm PBH}\equiv \Omega_{\rm PBH}/\Omega_{\rm DM} \sim 10^{-3}$ and $M_{\rm PBH}\sim 1\ M_\odot$ then $\sigma_H \sim 0.1$ \cite{Green:2021PBH}. In this case $\bar{a}_* \simeq 0.63$, with the majority of PBHs laying above this average~\cite{Harada:2017Spins}. 

{\it Let us now consider case (iii)}.   A long-range scalar-mediated force can facilitate growth of halos composed of heavy particles~\cite{Amendola:2017xhl,Savastano:2019zpr,Flores:2020drq,Domenech:2021uyx}. In addition, these same forces provide dissipation channels leading to PBH formation~\cite{Flores:2020drq}. In its minimal realization, this mechanism requires only one species of heavy particles interacting via Yukawa forces mediated by a light-scalar field. In this scenario, we allowed an asymmetry $\eta_\psi$ to develop in analogy to the baryon asymmetry $\eta_{\rm B}$. As with the standard model sector at high temperatures, sphaleron transitions also could occur in the dark sector. This could generate an asymmetry similar in scope to the baryon asymmetry.

Unlike gravitational instabilities, sufficiently strong scalar forces can lead to growth of perturbations  during the radiation dominated era 
\cite{Gradwohl:1992ue,Gubser:2004uh,Nusser:2004qu,Amendola:2017xhl,Savastano:2019zpr,Domenech:2021uyx}. The characteristic timescale for these forces is significantly shorter than the Hubble time, implying rapid structure formation. In the absence of energy dissipation, these structures would remain as virialized dark matter halos~\cite{Savastano:2019zpr}. However, these same long-range forces cause any accelerating charges to radiate scalar waves~\cite{Flores:2020drq}. The emission of scalar radiation removes energy from the virialized halos, leading to the formation of PBHs.

Let us examine spin distribution of the black holes generated by the ``fifth force" mechanism. 

To do so, we consider a light fermion $\psi$ interacting with a scalar field $\chi$:
\begin{equation}
\mathcal{L} \supset \frac{1}{2}m_\chi^2\chi^2 + m_\psi\bar{\psi} \psi - y\chi\bar{\psi}\psi + \cdots
.
\end{equation}

The $\chi$ field is assumed either very light, or massless $m_\chi \ll m_\psi^2/M_{\rm Pl}$ where $M_{\rm Pl} \approx 2\times 10^{18}$ GeV is the reduced Planck mass. The $\psi$ particles are either stable, or have a decay total decay width $\Gamma_\psi\ll m_\psi/M_{\rm Pl}$. This requirement ensures that there is a cosmological epoch where the $\psi$ particles are both nonrelativistic and decouple from equilibrium which is necessary for structure to form.

The strength of the Yukawa interaction, characterized by the parameter $\beta\equiv y(M_{\rm Pl}/m_\psi)\gg 1$ allows for the formation of structure in the linear regime during the radiation dominated era. Initially, a halo of $\psi$ particles may be subject to radiation pressure due to scalar interactions. This outward pressure becomes unimportant once the mean-free path of the $\chi$ particle exceeds the radius of the $\psi$ halo around $T_g\sim m_\psi/\ln(y^4M_{\rm Pl}/m_\psi)$. This temperature is close to the freeze-out temperature where the annihilation reactions $\bar{\psi}\psi \to \chi\chi$ freeze out which, for $y\sim 1$ and asymmetry $\eta_\psi \ll 1$, occurs when $T_f\sim m_\psi/36$~\cite{Graesser:2011wi}. In the limit of $\beta\gg 1$, the perturbations $\Delta(x,t) = \Delta n_\psi/n_\psi$ grow exponentially with a characteristic timescale $\tau_\Delta\equiv\Delta/(d\Delta/d t)$ smaller than the Hubble time. This implies rapid structure formation, with structures forming quickly after the $\psi$ particles decouple.

Eventually, the perturbations become nonlinear and the $\psi$ halos virialize. Before dissipation becomes important, interactions between neighboring halos will induce rotation. Motivated by the analogous quantity for gravity, we define the dimensionless spin parameter~\cite{Peebles:1971Rotation},
\begin{equation}
\label{eq:LamSpinParam}
\lambda_y\equiv \frac{J m_\psi}{yM_h^{3/2}R_h^{1/2}}
\end{equation}
where $J$ is the total halo angular momentum, and $M_h$ and $R_h$ are halo mass and radius respectively. Given the complexity of the dynamics involved, the true distribution for $\lambda_y$ should be determined using $N$-body simulations. In absence of this, we will assume that the initial distribution for $\lambda_y$ is functionally the same as the case with gravity. Numerical simulations have shown that the spin parameter distribution is well fit by a log-normal distribution~\cite{Efstathiou:1979Rotation},
\begin{equation}
p(\lambda_y)
d\lambda_y
=
\frac{1}{\sqrt{2\pi}\sigma_{\ln\lambda}}
\exp
\left[
-\frac{\ln^2(\lambda_y/\bar{\lambda}_y)}{2\sigma_{\ln\lambda}}
\right]\frac{d\lambda_y}{\lambda_y}
\end{equation}
where $\bar{\lambda}_y$ and $\sigma_{\ln\lambda}$ are parameters to be determined by numerical work.

When dissipation becomes important, the $\psi$ halos begin to contract as energy is removed via scalar radiation. There are several   dissipation channels which lead to the collapse of a given halo.
First, the motion could be coherent. Such a system emits scalar waves with a single frequency $\omega$ producing dipole radiation with $P_{\rm coh}\propto y^2N^2$. Second, the motion could be incoherent. In this case, each charge is treated as an individual source of radiation leading to a total power of $P_{\rm incoh}\propto y^2\omega^4R^2N$. Third, there is scalar bremsstrahlung radiation similar to free-free emission from plasma. In particular, we are interested in pairwise interactions of charges similar to the $e$ - $e$ component of free-free emission from plasma. Lastly, the contracting halo will become opaque and radiation will be trapped. In this case, radiation will only be allowed to escape from the surface and $P_{\rm surf}\sim 4\pi R^2 T_{\rm halo}^4$. The relevant timescale for the cooling of the halo is given by
\begin{equation}
\tau_{\rm cool}\equiv
\frac{E}{P_{\rm incoh} + P_{\rm coh} + P_{\rm brem} + P_{\rm surf} + \cdots},
\end{equation}
where $E = y^2N^2/R$. For a halo of a given radius, radiative cooling will become important once $\tau_{\rm cool} < H^{-1}$. Once cooling commences, the halo will quickly collapse and form a black hole with mass $M_{\rm BH}\sim M_h$.

In addition to carrying away energy from the virialized dark matter halos, scalar radiation can also remove angular momentum. In order understand the distribution of PBH spins, it is important that we track the evolution of the angular momentum as halos collapse. 

First, we will consider dissipation channels where the motion is oscillatory, namely coherent and incoherent radiation. A charge distribution of the form $\rho(x) = \rho(\mathbf{r}|\omega)\exp(i\omega t)$ suggests the decomposition $\chi(x) = u(\mathbf{r}|\omega)\exp(i\omega t)$. We can expand the solution for the amplitude for outgoing waves in the following manner:
\begin{equation}
u(\mathbf{r}|\omega)
=
\sum_{\ell = 0}^{\infty}
\sum_{m = -\ell}^{\ell}
\Lambda_{\ell m}(k)
h_\ell^{(1)}(kr)Y_\ell^m(\Omega)
,
\end{equation}
where $\omega/k = 1$, $h_\ell^{(1)}(x)$ are spherical Hankel functions of the first kind and $Y_{\ell m}(\Omega)$ are the spherical harmonics. The expansion coefficients are given by the integral relation,
\begin{equation}
\label{eq:ExpanCoeff}
\Lambda_{\ell m}(k)
=
iky
\int d^3r\ j_\ell(kr) Y_{\ell m}^*(\Omega)\rho(\mathbf{r}|\omega)
,
\end{equation}
where $j_\ell(x)$ are the spherical Bessel functions of the first kind.

The energy loss for a given spherical mode $(\ell,m)$ is given by
\begin{equation}
\label{eq:OscdEdt}
\frac{dE_{\ell m}}{dt}
=
\frac{1}{2}|\Lambda_{\ell m}(k)|^2
.
\end{equation}
It should be noted that in electromagnetism, $dE_{\rm EM}/dt\propto \ell(\ell + 1)$ implying that there is no $\ell = 0$ radiation for electromagnetic waves. The angular momentum loss due to scalar radiation with an oscillatory source is determined by
\begin{equation}
\label{eq:OscdJdt}
\frac{dJ_{\ell m}}{dt}
=
\frac{m}{2\omega}|\Lambda_{\ell m}(k)|^2
=
\frac{m}{\omega}
\left(
\frac{dE_{\ell m}}{dt}
\right),
\end{equation}
where $m = -\ell,\ldots, \ell$ as usual. Together \eqref{eq:OscdEdt} and \eqref{eq:OscdJdt} illustrate that for the $\ell = 0$ mode, energy will be taken away from the halo without the removal of angular momentum. In particular, for large wavelengths $kR\ll 1$ an order of magnitude estimate of the expansion coefficients~\eqref{eq:ExpanCoeff} gives
\begin{equation}
\Lambda_{\ell m}(k)
\sim ikyQ_{\rm tot}(kR)^{\ell}
\end{equation}
which demonstrates that the $\ell = 0$ mode dominates in the long-wavelength limit. In this circumstance, collapse of the halo will commence leaving the angular momentum unchanged.

Eventually, oscillatory dissipation channels  fall out of favor and cooling continues via nonoscillatory pathways. The nonoscillatory dissipation channels also carry angular momentum away from the collapsing halos. Both scalar bremsstrahlung and surface radiation are emitted isotropically from the dark matter halo when viewed in the corotating frame of the halo. However, in the lab frame the scalar quanta are blueshifted (redshifted) when emitted in a direction parallel  (antiparallel) to the motion of the halo. The rate of angular momentum loss in this case is 
\begin{equation}
\label{eq:NonOscdJdt}
\frac{dJ}{dt}
= -R\frac{dE}{dt}f(v)
\end{equation}
where
\begin{align}
f(v)
&\equiv
\frac{3\pi^2}{4}
\left[
\frac{(1 + v^2)\tanh^{-1}v - v}{v^2}
\right]\\[0.25cm]
&\approx
\pi^2
\left(v + \frac{2}{5}v^3 + \cdots \right)
\end{align}
and $v$ is the halo rotation velocity. Regardless of the solution to \eqref{eq:NonOscdJdt} we wish to determine if the angular momentum is emitted on a timescale comparable to the cooling timescale. To do so, we define 
$\tau_{J}^{-1}\equiv J^{-1} dJ/dt$. For $J\simeq M_h v R$ we find that
\begin{equation}
\frac{\tau_{\rm cool}}{\tau_J}
=
\mathcal{A}\left(\frac{R_0}{R}\right)
\frac{f(v)}{v}
\end{equation}
where $\mathcal{A}$ is defined by
\begin{equation}
\mathcal{A}
\equiv
\frac{y^2M_h}{m_\psi^2R_0}
=
\frac{M_h}{R_0}
\left(
\frac{\beta}{M_{\rm Pl}}
\right)^2
\end{equation}
and $R_0$ is the initial halo radius. For the benchmark parameters of Ref.~\cite{Flores:2020drq}, $y\sim 1$, $m_\psi\sim 1$ GeV, $M_h\sim 10^{-10}$ $M_\odot$ and $R_0\sim H^{-1}(T_g)\sim 10^{18}$ GeV$^{-1}$ we find that $\mathcal{A}\sim 10^{29}\gg 1$.

Therefore, for the relevant ranges of parameters~\cite{Flores:2020drq}, the timescale associated with angular momentum loss is significantly smaller than the energy-loss timescale and the Hubble time. Thus, angular momentum is removed from the system very efficiently, even in the case where the rotation velocity is nonrelativistic. Solving \eqref{eq:NonOscdJdt} for $v\ll 1$ and $R\ll R_0$ yields
\begin{equation}
\label{eq:AngMomEvo}
J(R)
=
J_0
\left(
\frac{R}{R_0}
\right)
\exp
\left[
-\mathcal{A}
\left(
\frac{R_0}{R}
\right)
\right]
.
\end{equation}
This result holds even for modest values of $\beta$.

For the parameters considered in \cite{Flores:2020drq}, bremsstrahlung was the main emission channel which was later followed by radiative cooling from the surface once the scalar radiation becomes trapped. As a halo collapses, its angular momentum evolves in accordance with Eq.~\eqref{eq:AngMomEvo}, resulting in a rapid spin-down. The swiftness of angular momentum loss is most clearly illustrated by the fact that $\tau_J\ll \tau_{\rm cool} < H^{-1}$. The quick removal of angular momentum results in a PBH with negligible spins at the time of formation.

The evolution of PBH spins to the present day  depends on the merger history and the details of accretion. To start, we will consider the spin evolution of BH distributions accessible by present-day gravitational wave experiments, i.e., for masses $\gtrsim 10\ M_{\odot}$. In this scenario, the bulk of dark matter is not PBHs as illustrated by present-day constraints on the  PBH-DM fraction, $f_{ \rm PBH}$~\cite{Carr:2020gox}.  These PBHs can seed the growth of dark matter halos shortly after the matter-radiation equality. The enhanced gravitational potential of this newly formed halo increases the accretion rate, dramatically effecting the final mass and spin of the seed black hole. The dimensionless Bondi-Hoyle accretion rate for $f_{\rm PBH} < 1$ is given by~\cite{Ricotti:2007au}
\begin{equation}
\begin{split}
\dot{m}(
f_{\rm PBH} < 1
)
\equiv 
\frac{\dot{M}_b}{\dot{M}_{\rm Ed}}
&=
(0.016\lambda)
\left(
\frac{1 + z}{1000}
\right)\times \\[0.25cm]
&\times
\left(
\frac{M_{\rm PBH}}{1\ M_\odot}
\right)\left(
\frac{v_{\rm eff}}{5.47\ \text{km s}^{-1}}
\right)^{-3}
\end{split}
\end{equation}
where $\lambda$ is the accretion eigenvalue for an isothermal gas, $v_{\rm eff}\equiv \sqrt{v_{\rm rel}^2 + c_s^2}$ with $v_{\rm rel}$ being the relative velocity of the PBH in question.

The relative velocity between PBHs and baryons for the era relevant to accretion is not well understood. The evolution of slow spinning PBHs was examined in Ref.~\cite{DeLuca:2020Evo} under the assumption that $v_{\rm eff}\sim c_s$. Until redshifts $z\lesssim 100$, the characteristic timescale associated with accretion exceeds the age of the Universe. Once growth does start, it proceeds until $z\sim 10$ \cite{DeLuca:2020Evo, Ricotti:2007au}. Present-day PBH masses $\lesssim 30\ M_\odot$ are expected to remain nonspinning for all redshifts whereas larger black holes will be near extremal up to redshifts $z\sim 10$. This conclusion does not account for supersonic motion at small scales discussed in Ref. \cite{Tsel:2010RelVel}, which could significantly suppress accretion.

The evolution of PBHs which may constitute dark matter is dramatically different. The fraction of PBHs that have undergone a merging event before some time $t$ is given by~\cite{Liu:2019Effects}
\begin{equation}
P_{\rm PBH}^{(1)}(t)
=
1.34\times 10^{-2}
\left(
\frac{M_*}{M_\odot}
\right)^{\frac{5}{37}}
\left(
\frac{t}{t_0}
\right)^{\frac{3}{37}}
f_{\rm PBH}^{\frac{16}{37}}\Upsilon_1,
\end{equation}
where $M_*$ is the characteristic mass scale associated with the mass function, and $\Upsilon_1$ is a dimensionless constant dependent on the form of mass function under consideration. For a Press-Schechter mass function defined over the dark matter window, as used in \cite{Flores:2020drq}, $\Upsilon_1\lesssim \mathcal{O}(10)$. Considering that $M_*\lesssim 10^{-10}\ M_\odot$ for PBHs relevant to explaining dark matter, we conclude that mergers are unimportant to the evolution of both mass and spin in this regime. 

In addition to this, we expect that accretion will be heavily suppressed. The dimensionless Bondi-Hoyle accretion rate (for $f_{\rm PBH} = 1$) is given by \cite{Ricotti:2007au}
\begin{equation}
\begin{split}
\dot{m}(f_{\rm PBH }= 1)
\equiv 
\frac{\dot{M}_b}{\dot{M}_{\rm Ed}}
=
&(1.8\times 10^{-3}\ \lambda)
\left(
\frac{1 + z}{1000}
\right)^3
\times\\[0.25cm]
&\times
\left(
\frac{M_{\rm PBH}}{1\ M_\odot}
\right)\left(
\frac{v_{\rm eff}}{5.47\ \text{km s}^{-1}}
\right)^{-3}
\end{split}
\end{equation}

Super-Eddinton accretion,  $\dot{m}\gtrsim 1$, is required for a significant spin-up. For sufficiently small black holes $\dot{m}\ll 1$, indicating that accretion is heavily suppressed. The combination of these facts points to a population of black holes which are largely unchanged from those at formation.

In summary, PBH spins reflect their cosmological origin.  Black holes that form from collapse of horizon-size density perturbations have negligible spins.  The black holes formed from merger of particles or scalar field solitons in the absence of radiative cooling can have a range of spins, from small to large.  However, in the presence of radiative cooling (which is essential in some formation scenarios), the angular momentum is removed from a collapsing halo faster than the energy, leading to slowly rotating black holes.  The prospects for measuring the black hole spin distribution with gravitational waves and other observations~\cite{Apostolatos:1994mx,Garcia-Bellido:2020pwq,Natwariya:2021xki} open a new window on the early Universe cosmology if some of the black holes are confirmed to have primordial origin. 

\begin{acknowledgments}
We thank T. Harada, K. Kohri, M. Sasaki, and C. Yoo for helpful discussions. 
This work was supported by the U.S. Department of Energy (DOE) Grant No. DE-SC0009937. The work of A.K. was also supported by World Premier International Research Center Initiative (WPI), MEXT, Japan, and by Japan Society for the Promotion of Science (JSPS) KAKENHI Grant No. JP20H05853. 
M.M.F. was supported by donors to the UCLA Department of Physics \& Astronomy.

\end{acknowledgments}


\bibliography{bibliography}
 
\end{document}